\documentstyle[12pt]{article}
\begin{document}
\bibliographystyle{unsrt} 
\vspace{10mm}

\begin{center}
{\Large Unification of Spacetime Symmetries of \\[2mm]
Massive and Massless Particles\footnote{Presented at the Second
German-Polish Symposium on New Ideas in the Theory of Fundamental
Interactions (Zakopane, September 11-15, 1995)}}
\\[3mm]
Y. S. Kim\\
Department of Physics, University of Maryland \\
College Park, Maryland 20742, U.S.A.
\end{center}

\begin{abstract}
The internal space-time symmetries of relativistic particles are
dictated by Wigner's little groups.  The $O(3)$-like little group
for a massive particle at rest and the $E(2)$-like little group of
a massless particle are two different manifestations of the same
covariant little group.  Likewise, the quark model and parton
pictures are two different manifestations of the one covariant
entity.
\end{abstract}

\section{Introduction}\label{sec1}
Eugene Wigner's 1939 paper on the Poincar\'e group is regarded as one
of the most fundamental papers in modern physics~\cite{wig39}.
Wigner observed there that relativistic particles have their internal
space-time degrees of freedom, and formulated their symmetries in terms
of the little groups of the Poincar\'e group.  He then showed that the
little groups for massive and massless particles are isomorphic to the
$O(3)$ and $E(2)$ groups respectively.

The purpose of this report is to emphasize that the little group is a
Lorentz-covariant entity and unifies the internal space-time symmetries
of both massive and massless particles, just as Einstein's $E = mc^{2}$
does for the energy-momentum relation.  On the other hand, Wigner did
not reach this conclusion in 1939, but his paper raised the following
questions.

\begin{itemize}
\item[1]
Like the three-dimensional rotation group, $E(2)$ is a three-parameter
group.  It contains two translational degrees of freedom in addition to
the rotation.  What physics is associated with the translational-like
degrees of freedom for the case of the $E(2)$-like little group?

\item[2]
As is shown by Inonu and Wigner~\cite{inonu53}, the rotation group $O(3)$
can be contracted to $E(2)$.  Does this mean that the $O(3)$-like little
group can become the $E(2)$-like little group in a certain limit?

\item[3]
It is possible to interpret the Dirac equation in terms of Wigner's
representation theory~\cite{raczka86}.  Then, why is it not possible
to find a place for Maxwell's equations in the same theory?

\item[4]
The proton had been known to have a finite space-time extension, and
it is believed to be a bound state of quarks.  Is it then possible to
construct a representation of the Poincar\'e group for particles with
space-time extensions?
\end{itemize}

As for the first question, it has been shown by various authors that the
translation-like degrees of freedom in the $E(2)$-like little group is
the gauge degree of freedom for massless particles~\cite{janner71,knp86}.
The second question will be addressed in detail in Sec.~\ref{sec2}.  As
for the third question, Weinberg found a place for the gauge-invariant
electromagnetic fields in the Wigner formalism by constructing from the
SL(2,c) spinors all the representations of massless fields which are
invariant under gauge transformations~\cite{wein64}.  It has also been
shown that gauge-dependent four-potentials can also be constructed within
the SL(2,c) framework~\cite{hks86}.  The Maxwell theory and the Poincar\'e
group are now perfectly consistent with each other.

The fourth question is about whether Wigner's little groups are
applicable to high-energy particle physics where accelerators produce
Lorentz-boosted extended hadrons such as high-energy protons.  The
question is whether it is possible to construct a representation of the
Poincar\'e group for hadrons which are believed to be bound states of
quarks~\cite{knp86,fkr71}.  This representation should describe
Lorentz-boosted hadrons.  Next question then is whether those boosted
hadrons give a description of Feynman's parton picture~\cite{fey69}
in the limit of large momentum/mass~\cite{knp86,kn77a}.  We shall
concentrate on this fourth question in this report.

\section{Little Groups of the Poincar\'e Group}\label{sec2}
The little group is the maximal subgroup of the Lorentz group which
leaves the four-momentum invariant.  While leaving the four-momentum
invariant, the little group governs the internal space-time symmetries
of relativistic particles.  The Lorentz group is generated by three
rotation generators $J_{i}$ and three boost generators $K_{i}$.  If a 
massive particle is at rest, the little group is the three-dimensional
rotation group generated by $J_{1}$, $J_{2}$ and $J_{3}$.  The
four-momentum is not affected by this rotation, but the spin variable
changes its direction.  For a massless particle moving along the $z$
direction, Wigner observed that the little group is generated by
$J_{3}, N_{1}$ and $N_{2}$, where
\begin{equation}\label{e2gen}
N_{1} = J_{1} + K_{2} , \qquad  N_{2} = J_{2} - K_{1} , 
\end{equation}
and that these generators satisfy the Lie algebra for the two-dimensional 
Euclidean group.  Here, $J_{3}$ is like the rotation generator, while 
$N_{1}$ and $N_{2}$ are like translation generators in the two-dimensional 
Euclidean plane.

In 1953, Inonu and Wigner formulated this problem as the contraction of
$O(3)$ to $E(2)$.  How about then the little groups which are isomorphic
to $O(3)$ and $E(2)$?  It is reasonable to expect that the $E(2)$-like
little group be obtained as a limiting case for of the $O(3)$-like little
group for massless particles~\cite{misra76}.  It is shown that, under
the boost along the $z$ direction, the rotation generator around the $z$
axis remains invariant, but the transverse rotation generators $J_{1}$
and $J_{2}$ become generators of gauge transformations in the large-boost
limit~\cite{hks83pl}.  It was later shown that the little group for
massless particles has the geometry of the cylindrical
group~\cite{kiwi87jm}

In the following sections, we shall discuss how the concept of little
groups are applicable to space-time symmetries of relativistic extended
hadrons.

\section{Covariant Harmonic Oscillators}\label{sec3}
Let us consider a hadron consisting of two quarks.  If the space-time 
position of two quarks are specified by $x_{a}$ and $x_{b}$ respectively, 
the system can be described by the variables 
\begin{equation}
X = (x_{a} + x_{b})/2 , \qquad x = (x_{a} - x_{b})/2\sqrt{2} .
\end{equation}
The four-vector $X$ specifies where the hadron is located in space and
time, while the variable $x$ measures the space-time separation between
the quarks.

The portion of the wave function which is subject to the $O(3)$-like
little group takes the form
\begin{equation}\label{2.6}
\psi^{n}_{0}(z,t) = \left({1\over \pi n! 2^{n}}\right)^{1/2} H_{n}(z)
\exp \left\{-{1\over 2}\left(z^{2} + t^{2} \right) \right\} .
\end{equation}
The subscript 0 means that the wave function is for the hadron at rest.
The above expression is not Lorentz-invariant, and its localization
undergoes a Lorentz squeeze as the hadron moves along the $z$
direction~\cite{knp86}.

It is convenient to use the light-cone variables to describe Lorentz
boosts.  The light-cone coordinate variables are
\begin{equation}
u = (z + t)/\sqrt{2} , \qquad v = (z - t)/\sqrt{2} .
\end{equation}
In terms of these variables, the Lorentz boost takes the simple form
\begin{equation}
u' = e^{\eta } u , \qquad v' = e^{-\eta } v ,
\end{equation}
where $\eta $ is the boost parameter and is $\tanh ^{-1}(v/c)$.  This is
a ``squeeze'' transformation.

In Eq.(\ref{2.6}), the localization property of the wave function is
determined by the Gaussian factor, and we shall therefore study the
ground state and its wave function
\begin{equation}\label{wf0}
\psi_{0}(z,t) = \left({1 \over \pi} \right)^{1/2}
\exp \left\{-{1\over 2} (u^{2} + v^{2}) \right\} . 
\end{equation}
If the system is boosted, the wave function becomes
\begin{equation}\label{wf1}
\psi_{\eta }(z,t) = \left({1 \over \pi }\right)^{1/2}
\exp \left\{-{1\over 2}\left(e^{-2\eta }u^{2} + 
e^{2\eta }v^{2}\right)\right\} .
\end{equation}
The wave function of Eq.(\ref{wf0}) is distributed
within a circular region in the $u v$ plane, and thus in the $z t$ plane.  
On the other hand, the wave function of Eq.(\ref{wf1}) is distributed in
an elliptic region whose major and minor axes are along the light-cone
axes.  The wave function becomes Lorentz-squeezed!

\section{Feynman's Parton Picture}\label{sec4}

In order to explain the scaling behavior in inelastic scattering,
Feynman in 1969 observed that a fast-moving hadron can be regarded as
a collection of many ``partons'' whose properties do not appear to be
identical to those of quarks \cite{fey69}.  For example, the number of
quarks inside a static proton is three, while the number of partons in a
rapidly moving proton appears to be infinite.  The question then is how
the proton looking like a bound state of quarks to one observer can
appear different to an observer in a different Lorentz frame?  Feynman
made the following systematic observations.

\begin{itemize}

\item[a] The picture is valid only for hadrons moving with velocity
       close to that of light.

\item[b] The interaction time between the quarks becomes dilated, and
        partons behave as free independent particles.

\item[c] The momentum distribution of partons becomes widespread as the
       hadron moves fast.

\item[d] The number of partons seems to be infinite or much larger than
      that of quarks.

\end{itemize}

\noindent Because the hadron is believed to be a bound state of two or
three quarks, each of the above phenomena appears as a paradox,
particularly b) and c) together.  In order to resolve this paradox, we
need a momentum-energy wave function.  If the quarks have the
four-momenta $p_{a}$ and $p_{b}$, we can construct two independent
four-momentum variables~\cite{fkr71}
\begin{equation}
P = p_{a} + p_{b} , \qquad q = \sqrt{2}(p_{a} - p_{b}) .
\end{equation}
The four-momentum $P$ is the total four-momentum and is thus the hadronic 
four-momentum, while $q$ measures the four-momentum separation between
the quarks.  In the light-cone coordinate system, the momentum-energy
variables are
\begin{equation}
q_{u} = (q_{0} - q_{z})/\sqrt{2} ,  \qquad
q_{v} = (q_{0} + q_{z})/\sqrt{2} .
\end{equation}
Then the momentum-energy wave function takes the form
\begin{equation}\label{wf2}
\phi_{\eta }(q_{z},q_{0}) = \left({1 \over \pi }\right)^{1/2} 
\exp\left\{-{1\over 2}\left(e^{-2\eta}q_{u}^{2} + 
e^{2\eta}q_{v}^{2}\right)\right\} .
\end{equation}

The momentum wave function is also squeezed, and the parton momentum
distribution becomes wide-spread as the hadronic speed approaches the
speed of light.  It is thus possible to calculate the parton
distribution by boosting a hadronic wave function in the rest frame.
The calculation based on this oscillator model gives a reasonable
agreement with the measured parton distribution function~\cite{hussar81}.

Let us go back to the Lorentz-squeezed space-time wave function given in
Eq.(\ref{wf1}).  This wave function gives two time intervals corresponding
to the major and minor axes of the elliptic distribution.  As the hadronic
speed approaches the speed of light, the major axis corresponds to the
period of oscillation, and it increases by factor of $e^{\eta}$.  This
period measures the interaction time among the quarks.

The external signal comes into the hadron in the direction opposite to
the hadron momentum.  Thus the minor axis of the ellipse measures the
the time the external signal spends inside the hadron.  This is the
interaction time between one of the quark and the external signal. This
time interval decreases as $e^{-\eta}$.  The ratio of the interaction
time to the oscillator period becomes $e^{-2\eta}$.  The energy of each
proton coming out of the Fermilab accelerator is 900 GeV.  This leads
the ratio to $10^{-6}$, which is indeed a small number.  The external
signal is not able to sense the interaction of the quarks among
themselves inside the hadron.  This is why partons appear as free
particles with a wide-spread momentum distribution.

The internal space-time symmetry of hadrons in the quark model can be
framed into the $O(3)$-like little group when they are slowly moving
particles.  It is also possible to frame the symmetry of the parton model
into the $E(2)$-like little group for massless particles~\cite{kim89}.
It is indeed gratifying to note that these two seemingly different
symmetries are two different manifestations of the same covariant
symmetry.

\end{document}